\documentclass[12pt]{article}
\usepackage{graphicx}
\usepackage{subfigure}
\usepackage{amssymb}
\usepackage{amsfonts}
\usepackage{latexsym}

\usepackage[usenames]{color}
\usepackage{latexsym}
\usepackage{epstopdf} 
\usepackage[normalem]{ulem}

\input epsf.tex

\newcommand{\beq}{\begin{equation}}
\newcommand{\eeq}{\end{equation}}
\newcommand{\be}{\begin{equation}}
\newcommand{\ee}{\end{equation}}
\newcommand{\bea}{\begin{eqnarray}}
\newcommand{\eea}{\end{eqnarray}}

\makeatletter
\renewcommand{\theequation}{\thesection.\arabic{equation}}
\@addtoreset{equation}{section}
\makeatother

\def\href#1#2{#2}

\usepackage{amsmath}	
\begin{document}

\baselineskip=15.5pt
\pagestyle{plain}
\setcounter{page}{1}

\begin{titlepage}
\begin{flushleft}
       \hfill                       
\end{flushleft}

\begin{center}
  {\huge Andreev reflection \\   
  \vspace*{2mm}
  at
Hadron/Color superconductor interface 
\vspace*{2mm}
}
\end{center}

\begin{center}

\vspace*{5mm}
{\large 
Yuki Juzaki\footnote[1]{\tt yuki.juzaki@gmail.com} and
Motoi Tachibana\footnote[2]{\tt motoi@cc.saga-u.ac.jp}}\\
\vspace*{2mm}
{
Department of Physics, Saga University, Saga 840-8502, Japan}
\vspace*{2mm}
\end{center}

\begin{center}
{\large Abstract}
\end{center}
We consider the phenomenon of the Andreev reflection of "hadrons" at the interface between hadronic and color superconducting phases, which are expected to appear in the neutron star interior. Here, hadrons are defined as a superposition of constituent quarks, each of which is Andreev-reflected. 
We study what kind of reflections are possible to come out of incident mesons and baryons in the hadronic phase, attached to different color superconducting phases.
Then, some peculiar patterns of the reflections are obtained. 

\noindent

\begin{flushleft}

\end{flushleft}
\end{titlepage}

\vspace{1cm}
\section{Introduction}




It has been anticipated that matter at high baryon number density (or equivalently large baryon chemical
potential) and/or high temperature reveals various interesting properties, such as quark-gluon plasma (QGP), superfluidity of neutrons, superconductivity of protons, Bose-Einstein condensate of mesons and superconductivity of quarks which is called color superconductivity (CSC) \cite{CSC}. It is believed that those states of matter can be described by Quantum Chromodynamics (QCD), i.e., dynamics of quarks (constituents of hadrons) and gluons (mediator of the interaction between quarks). It is one of the major goals in hadron physics to comprehend what kind of matters exists at different temperatures and chemical potentials. This is nothing but the question of the phase diagram of QCD \cite{review}.

Elucidating the QCD phase diagram is also important from experimental and observational points of view \cite{YHM}. QCD at high temperature is intimately connected with physics in the early universe right after Big-Bang, while QCD at high baryon density (and relatively small temperature) is related to physics of compact stars such as neutron stars, hypothetical quark stars and even black-holes. Quite recently, compact stars have been remarkable objects since their binary systems are widely recognized as a source of the gravitational wave signal \cite{LIGO}.

Neutron star is supposed to have an onion-like structure with various kinds of phases (for a review of neutron stars, see \cite{NGK, Shapiro}). This means that there exist interfaces consisting of different phases in the neutron star interior. One of the authors studied the particle scattering happening at such interfaces before \cite{ST} and found an interesting phenomena called Andreev reflection, which was originally discovered in the context of condensed matter physics \cite{AR}. The Andreev reflection is a peculiar reflection at the metal-superconductor interface. An incident electron in metal hits the interface, then a hole, which carries information of the superconductor, is reflected. This is something like when we throw a ball against a wall, the reflected ball is not the original one, but changed with having information behind the wall. This phenomena is intuitively interpreted as follows. When the incident electron hits the interface, a part of its component goes into superconductor in the sense of quantum mechanics, namely, the wave function. If the incident electron energy is lower than the superconducting gap, the incident electron itself is not allowed to penetrate, but possible to pass through in the form of the Cooper pair. Then, the incident electron accompanies another electron below the Fermi level. Consequently, a hole is reflected.

The purpose of this paper is to extend our previous study \cite{ST} to the case with hadron-color superconductor (CSC) interface. In section 2, we briefly review our previous study. In section 3, we study the Andreev reflection at hadron/CSC interface in a systematic way. Section 4 is devoted to summary and discussions.

\section{Andreev reflection in color superconductor}


The effective hamiltonian describing quark interaction with the diquark condensate (i.e., quark Cooper pairing) 
can be written at the mean field level \cite{ST}:
\begin{eqnarray}
H&=& \int d^3 x \left [ \sum_{a, i}\psi^{i\dagger}_a(-i\vec{\alpha}\cdot \nabla-\mu)\psi^i_a
+\sum_{a,b,i,j}\Delta^{ij}_{ab}\left (\psi^{iT}_aC\gamma_5 \psi^j_b \right )+{\rm h.c} \right ],
\label{2-1}
\end{eqnarray}
where $\psi^i_a$ is the Dirac spinor with $a, b$ and $i, j$ being color and flavor indices, respectively.
$C$ is the charge conjugation matrix and $\mu$ the quark chemical potential. $\Delta^{ij}_{ab}$ is the gap matrix.
For 2 flavor color superconductivity (2SC), it becomes
\begin{equation}
\Delta^{ij}_{ab}=\tilde{\Delta}\epsilon^{ij}\epsilon_{abB},
\label{2-2}
\end{equation}
where $i, j=u{\rm (up)},d{\rm (down)}$ and $a, b=R {\rm (red)},G {\rm (green)}$.
The third direction in color space is chosen as blue (denoted by $B$) here. For 3 flavor case, 
on the other hand, the gap matrix takes the form as
\begin{equation}
\Delta^{ij}_{ab}=\Delta\epsilon^{ijK}\epsilon_{abK}=\Delta(\delta^i_a \delta^b_j-\delta^i_b \delta^j_a),
\label{2-3}
\end{equation}
where $i, j=u{\rm (up)},d{\rm (down)}, s{\rm (strange)}$ and 
$a, b=R {\rm (red)},G {\rm (green)}, B{\rm (blue)}$. This is the color flavor-loked (CFL) condensate \cite{CFL}. 
The general form of the gap matrix can be written as follows:
\begin{eqnarray}
\Delta^{ij}_{ab} = 
\begin{pmatrix}
0 & \Delta_{ud} & \Delta_{us} & & & & & & \\
\Delta_{ud} & 0 & \Delta_{ds} & & & & & & \\
\Delta_{us} & \Delta_{ds} & 0 & & & & & & \\
& & & 0 & -\Delta_{ud}& & & & \\
& & & -\Delta_{ud} & 0 & & & & \\
& & & & & 0 & -\Delta_{us} & & \\
& & & & & -\Delta_{us} & 0 & & \\
& & & & & & & 0 & -\Delta_{ds} \\
& & & & & & & -\Delta_{ds} & 0 \\
\end{pmatrix}
\label{2-4}
\end{eqnarray}
in the basis
\begin{eqnarray}
\left( u_{R}, \ d_{G}, \ s_{B}, \ d_{R}, \ u_{G}, \ s_{R}, \ u_{B}, \ s_{G},\ d_{B} \right).
\label{2-5}
\end{eqnarray}
From this, one finds that for instance, a red up quark ($u_R$) is Andreev-reflected as a green down hole 
($d_G^H$) or a blue strange hole ($s_B^H$). The Andreev reflection of quarks is summarized in Table 1.

\begin{table}[h]
\centering
  \begin{tabular}{|c|c|} \hline
  incident quark   &  reflected hole \\ \hline \hline
              $u_{R}$ & $d_{G}^{H}$ or $s_{B}^{H}$   \\ \hline
              $u_{G}$ & $d_{R}^{H}$    \\ \hline
              $u_{B}$ &      $s_R^{H}$         \\ \hline 
              $d_{R}$ & $u_{G}^{H}$   \\ \hline
              $d_{G}$ & $u_{R}^{H}$ or $s_{B}^{H}$    \\ \hline
              $d_{B}$ &      $s_G^{H}$         \\ \hline 
              $s_{R}$ & $u_{B}^{H}$   \\ \hline
              $s_{G}$ & $d_{B}^{H}$    \\ \hline
              $s_{B}$ &      $u_{R}^{H}$ or $d_{G}^{H}$        \\ \hline 
  \end{tabular}
  \caption{Andreeev reflection of quarks}
\end{table}

In 2SC phase,
\begin{eqnarray}
&\Delta_{us}& = \Delta_{ds} = 0, \ \Delta_{ud} = \tilde{\Delta},
\label{2-6}
\end{eqnarray}
while in CFL phase,
\begin{eqnarray}
&\Delta_{us}& = \Delta_{ds} = \Delta_{ud} = \Delta.
\label{2-7}
\end{eqnarray}

The equations of motion obtained from (\ref{2-1}) are so called Bogoliubov - de-Gennes (BdG) equations
\begin{eqnarray}
\begin{pmatrix}
-i\vec{\alpha}\cdot \nabla - \mu & \Delta C\gamma_5 \\
\Delta C\gamma_5 & i\vec{\alpha}\cdot \nabla +\mu \\
\end{pmatrix}
\begin{pmatrix}
\psi_p \\
\psi_h  \\
\end{pmatrix}
= E
\begin{pmatrix}
\psi_p \\
\psi_h  \\
\end{pmatrix}.
\label{2-8}
\end{eqnarray}
The BdG equations with the vanishing gap can be interpreted as those in the free quark (FQ) phase while
with the non-vanishing one as the color superconducting (CSC) phase. 

Suppose here that there is a sharp interface between the FQ/CSC phases. Then the scattering of a single quark 
at the interface, which comes from the FQ phase, can be modeled as follows: \\
\\
(1) The interface is perpendicular to, say, the $x$ axis and is located at $x=0$. \\
(2) The gap is only a function of $x$ and takes the step function form $\Delta(x)=\Delta\Theta(x)$.\\
(3) Region $x<0$ describes the FQ phase while region $x>0$ the CSC phase.\\
(4) General solution of the BdG equations in each phase is obtained.\\
(5) Boundary conditions to match the solution in each phase at $x=0$ are imposed.\\
\\
As was stated in the previous section, our main interest in this paper is to extend the above study of a single-quark scattering into the multi-quark case, namely, the Andreev reflection between the hadronic and the CSC phases.


\section{ Andreev reflection at Hadron/CSC interface}
\label{sec:backreaction}


In the previous section, we reviewed the scattering problem of a single quark at the interface between
free quark and color superconducting phases. In this section, let us consider the same issue into 
the interface consisting of hadron and the CSC phases. The question is how the single-quark scattering
problem is extended to the multi-quark one. This is, in general, a very hard problem and so far we have
no clear answer. So let us make the following assumption: quarks in the hadronic phase are not interacting
with each other, but they be always in the color-singlet states. This is the constituent quark picture. Hadrons in here are defined as a superposition of every single quark. So if a hadron hits the hadron/CSC interface, each quark inside the hadron is Andreev-reflected and the only color-singlet combination is left out of the reflected quarks. 

In the case of hadron/2SC interface, the Andreev reflection of $s$ quark from the hadronic side does not occur because it cannot make any Cooper pairing in 2SC phase.
In the case of hadron/CFL interface, on the other hand, $s$ quark can make a Cooper pairing with other quarks
so that the Andreev reflection does occur. Moreover, a blue quark with any flavor cannot make any Cooper pairing in 2SC phase. With this in mind, we shall describe the Andreev reflection for different cases below. 

\subsection{Andreev reflection of mesons}

 

\subsubsection{Andreev reflection of mesons without $s$ quarks}

First of all, let us discuss the Andreev reflection of mesons without $s$ quarks.
As such an example, we have a charged pion, $\pi^{+}$, which consists  of a $u$ quark and 
an anti-$d$ ($\bar{d}$) quark. Meson is always a color-singlet superposition of 
red($R$)-anti-red($\bar{R}$), green($G$)-anti-green($\bar{G}$) and blue($B$)-anti-blue($\bar{B}$).
So we define here "meson" having a color component $a (=R, G, B)$ as follows:
\begin{eqnarray}
M_{a} \equiv i_{a}\bar{j}_{a}, \qquad {\rm (no \ sum)}
\end{eqnarray}
where $i$ and $j$ are flavor indices. Then, from the gap matrix structure (\ref{2-4}) and Table 1, 
one easily finds that
$\pi^{+}_{R}$ ($\pi^{+}_{G}$) is Andreev-reflected as a hole of $\pi^{-}_{G}$ ($\pi^{-}_{R}$), respectively:
\begin{eqnarray}
\pi^{+}_{R} = u_{R}\bar{d}_{R} \ \to \ (d_{G}\bar{u}_{G})^{H} \equiv (\pi^{-}_{G})^{H}, \\
\pi^{+}_{G} = u_{G}\bar{d}_{G} \ \to \ (d_{R}\bar{u}_{R})^{H} \equiv (\pi^{-}_{R})^{H}.
\end{eqnarray}
$\pi^{+}_{B}$, on the other hand, is not Andreev-reflected but reflected as $\pi^{+}_{B}$ itself. 
So the result is that $\pi^{+}$ is partially Andreev-reflected and partially ordinary reflected.
Then the question is how to interpret this. To this end, we remark that a hole is a sort of charge conjugation
of a particle and could be regarded as a anti-particle. Relying on the idea,
we suppose that $(\pi^{-}_{G})^{H}$ corresponds to $\pi^{+}_{G}$ so that $\pi^{+}$ is consequently reflected as $\pi^{+}$ at the hadron/2SC interface.

In the case of hadron/CFL interface, since $u_{B}$ and $\bar{d}_{B}$ also join the Cooper pairing, we expect $\pi^{+}_{B}$ is Andreev-reflected as well. However, this is not the case because the color-singlet state is not realized:
\begin{eqnarray}
\pi^{+}_{B} = u_{B} \bar{d}_{B} \ \to \ s_{R}\bar{s}_{G} = \times
\end{eqnarray}
Here $\times$ shows that the Andreev reflection does not occur. To summarize, mesons without $s$ quarks are reflected as the original ones in both 2SC and CFL cases. See Table 2.

\begin{table}[h]
\centering
  \begin{tabular}{|c|c|c|c|} \hline
  incident particle   &  incident(color)  & reflected(color) & reflected particle \\ \hline \hline
             & $\pi^{+}_{R}$ & $(\pi^{-}_{G})^{H}$  & \\ \cline{2-3}
   $\pi^{+}$ & $\pi^{+}_{G}$ & $(\pi^{-}_{R})^{H}$  & $\pi^{+}$ \\ \cline{2-3}
             & $\pi^{+}_{B}$ &      $\times$        & \\ \cline{2-3}\hline 
             & $\pi^{-}_{R}$ & $(\pi^{+}_{G})^{H}$  & \\ \cline{2-3}
   $\pi^{-}$ & $\pi^{-}_{G}$ & $(\pi^{+}_{R})^{H}$  & $\pi^{-}$ \\ \cline{2-3}
             & $\pi^{-}_{B}$ &      $\times$        & \\ \cline{2-3}\hline 
             & $\pi^{0}_{R}$ & $(\pi^{0}_{G})^{H}$  & \\ \cline{2-3}
   $\pi^{0}$ & $\pi^{0}_{G}$ & $(\pi^{0}_{R})^{H}$  & $\pi^{0}$ \\ \cline{2-3}
             & $\pi^{0}_{B}$ &      $\times$        & \\ \cline{2-3}\hline 
  \end{tabular}
  \caption{Andreeev reflection of pions}
\end{table}

\subsubsection{Andreev reflection of mesons with $s$ quarks}

Next, let us discuss the Andreev reflection of mesons with $s$ quarks. As such an example, we consider a charged K meson, $K^{+}=u\bar{s}$. In the case of hadron/2SC interface, the Andreev reflections for any color component do not occur and therefore $K^{+}$ is reflected as the original one.

In the case of hadron/CFL interface, we see the following results:
\begin{eqnarray}
K^{+}_{R} = u_{R}\bar{s}_{R} \ \to \ (s_{B} \bar{u}_{B})^{H} = (K^{-}_{B})^{H} \\
K^{+}_{B} = u_{B}\bar{s}_{B} \ \to \ (s_{R} \bar{u}_{R})^{H} = (K^{-}_{R})^{H}
\end{eqnarray}
On the other hand, for the green component of $K^{+}$, i.e., $K^{+}_{G}$, the Andreev reflection
does occur unlike the 2SC case, but it cannot be the color-singlet. Consequently, $K^{+}$ is reflected as $K^{+}$. Lastly, we considered $\eta$ meson as an incident particle in both 2SC and CFL cases. Then we found that the Andreev reflection for the incident $\eta$ meson does not occur for any color component. The results of this section is summarized in Table 3.


\begin{table}[h]
\centering
  \begin{tabular}{|c|c|c|c|} \hline
  incident particle   &  incident(color)  & reflected(color) & reflected particle \\ \hline \hline
             & $K^{+}_{R}$ & $(K^{-}_{B})^{H}$ & \\ \cline{2-3}
   $K^{+}$   & $K^{+}_{G}$ & $\times$     & $K^{+}$ \\ \cline{2-3}
             & $K^{+}_{B}$ & $(K^{-}_{R})^{H}$ & \\ \cline{2-3}\hline 
             & $K^{-}_{R}$ & $(K^{+}_{B})^{H}$ &\\ \cline{2-3}
   $K^{-}$   & $K^{-}_{G}$ & $\times$ & $K^{-}$ \\ \cline{2-3}
             & $K^{-}_{B}$ & $(K^{+}_{R})^{H}$ & \\ \cline{2-3}\hline 
             & $K^{0}_{R}$ & $\times$ & \\ \cline{2-3}
   $K^{0}$   & $K^{0}_{G}$ & $(\bar{K}^{0}_{B})^{H}$ & $K^{0}$ \\ \cline{2-3}
             & $K^{0}_{B}$ & $(\bar{K}^{0}_{G})^{H}$ & \\ \cline{2-3}\hline 
               & $\bar{K}^{0}_{R}$ & $\times$  & \\ \cline{2-3}
 $\bar{K}^{0}$ & $\bar{K}^{0}_{G}$ & $(K^{0}_{B})^{H}$ & $\bar{K}^{0}$ \\ \cline{2-3}
               & $\bar{K}^{0}_{B}$ & $(K^{0}_{G})^{H}$ & \\ \cline{2-3}\hline
  \end{tabular}
  \caption{Andreev reflection of K mesons}
\end{table}

\subsection{Andreev reflection of baryons}

Let us move on to arguing the Andreev reflection of baryons. Since any baryon includes one blue quark,
the Andreev reflection of baryons does not occur  at hadron/2SC interface, as described in the previous subsection. So we concentrate on the case of hadron/CFL interface below.

Let us first consider a neutron, $n=udd$ where two of three quarks are the same flavor ones. Then a neutron consists of a certain sum of six color-singlet combinations:
\begin{eqnarray}
u_{R}d_{G}d_{B}, \quad
u_{R}d_{B}d_{G}, \quad
u_{G}d_{R}d_{B}, \quad
u_{G}d_{B}d_{R}, \quad
u_{B}d_{R}d_{G}, \quad
u_{B}d_{G}d_{R} 
\nonumber
\end{eqnarray}
From the gap matrix structure (\ref{2-4}), one obtains the following results:
\begin{eqnarray}
u_{R}d_{G}d_{B} \to (s_{B}u_{R}s_{G})^H &=& (\Xi^{0})^{H},\\
u_{R}d_{B}d_{G} \to (s_{B}s_{G}u_{R})^H &=& (\Xi^{0})^{H}, \\
u_{G}d_{R}d_{B} \to (d_{R} s_{G}u_{G})^H &=& \ \times, \\
u_{G}d_{B}d_{R} \to (d_{R} s_{G}u_{G})^H &=& \ \times, \\
u_{B}d_{R}d_{G} \to (s_{R}u_{G}s_{B})^H &=& (\Xi^{0})^{H}, \\
u_{B}d_{G}d_{R} \to (s_{R}s_{B}u_{G})^H &=& (\Xi^{0})^{H}. 
\end{eqnarray}

As previously, $\times$ shows that the Andreev reflection does not occur. Therefore in this case, an incident neutron is reflected as a hole of $\Xi^0$ as well as neutron and its ratio is two to one. In Table 4, the results of baryons with 2 different flavors are summarized. Note here that unlike the case of mesons, a hole is not interpreted as an anti-particle. Also we are neglecting the effects of finite quark masses.


\begin{table}[h]
\centering
\begin{tabular}{|c|c|c|} \hline
  incident particle   &  constitution &  reflected particles \\ \hline \hline
  $p$          & $uud$ & $p$, \ $(\Xi^{-})^{H}$ \\ \cline{2-2} \hline
$n$          & $udd$ & $n$, \ $(\Xi^{0})^{H}$ \\ \cline{2-2} \hline
$\Delta^{+}$ & $uud$ & $\Delta^{+}$, \ $(\Xi^{-})^{H}$ \\ \cline{2-2} \hline
$\Delta^{0}$ & $udd$ & $\Delta^{0}$, \ $(\Xi^{0})^{H}$ \\ \hline
$\Sigma^{+}$ & $uus$ & $\Sigma^{+}$, \ $(\Sigma^{-})^{H}$ \\  \hline
$\Sigma^{-}$ & $dds$ & $\Sigma^{-}$, \ $(\Sigma^{+})^{H}$ \\  \hline
$\Xi^{0}$    & $uss$ & $\Xi^{0}$, \ $(\Delta^{0})^{H}$ \ or \ $(n)^{H}$ \\  \hline
$\Xi^{-}$    & $dss$ & $\Xi^{-}$, \ $(\Delta^{+})^{H}$ \ or \ $(p)^{H}$ \\  \hline
\end{tabular}
\caption{Baryons with two different flavors}
\end{table}

Next let us consider baryons with 3 different flavors such as $\Lambda=uds$. In this case,
\begin{eqnarray}
u_{R}d_{G}s_{B} &\to& uds = (\Lambda)^{H}, \\
u_{B}d_{R}s_{G} &\to& s_{R}u_{G}d_{B} = (\Lambda)^{H}, \\
u_{G}d_{B}s_{R} &\to& d_{R}s_{G}u_{B} = (\Lambda)^{H}, \\
u_{R}d_{B}s_{G} &\to& \ * \ s_{G}d_{B} = \ \times, \\
u_{G}d_{R}s_{B} &\to& d_{R}u_{G} \ * \ = \ \times, \\
u_{B}d_{G}s_{R} &\to& s_{R} * \ u_{B} = \ \times.
\end{eqnarray}
The reflected particles are $\Lambda$ itself and its own hole $(\Lambda)^{H}$. $*$ shows that there are two possible reflections due to the gap matrix structure. But in any case, we cannot achieve the color-singlet combinations. Table 5 summarizes  the results.

\begin{table}[h]
\centering
\begin{tabular}{|c|c|c|} \hline
  incident particle   &  constitution &  reflected particles \\ \hline \hline
  $\Lambda$ & $uds$ & $\Lambda$, \ $(\Lambda)^{H}$ \\ \cline{2-2} \hline
$\Sigma^{0}$ & $uds$ & $\Sigma^{0}$, \ $(\Sigma^{0})^{H}$ \\ \cline{2-2} \hline
\end{tabular}
\caption{Baryons with three different flavors}
\end{table}

Lastly, let us comment on the case of baryons with one flavor such as $\Delta^{++}=uuu$. In this case, it easily turns out that we cannot make any color-singlet combination. Therefore, the Andreev reflection does not occur.


\section{Summary and Discussions}

In this paper, we were interested in the Andreev reflection at the interface between hadronic
and color superconducting phases. Hadrons were defined as a superposition of constituent quarks, each of
which is described by the BdG equations. The reflected quarks (holes, indeed) must form the color-singlet states. Based on this observation, systematic studies were performed for the meson and baryon cases, respectively. Then we obtained some peculiar patterns of reflected hadrons. 

Here is some future perspective. In this study, we have not performed any concrete computations of quantities such as reflection and transmission probabilities as well as probability currents, which were calculated before \cite{ST}. These will be necessary when we try to apply the results obtained here into physics of neutron stars. Besides, we have not taken into account the influence of magnetic field to the BdG equations in this paper. This is also worth to considering.

Furthermore, in the original paper by Andreev \cite{AR}, the BdG equations provided a nice interpretation for excitations in different phases (electrons and holes in conductor while quasi-articles in superconductor). How such an interpretation works, however, is not guaranteed in the case of hadron/CSC interfaces. This is because the system we argued here is relativistic so that creation and annihilation of particles are possible. This suggests that we have to consider our problem at the field theoretical level. Then the previous work, where one of the authors joined, might give some insight \cite{HTYB}. This will be reported elsewhere in the future.





\section*{Acknowledgments}
The authors thank the Yukawa Institute for Theoretical Physics at Kyoto University. Discussions during the YITP workshop YITP-W-19-08 on "Thermal Quantum Field Theory and Their Applications" were useful to complete this work. They are also grateful M. Sadzikowski for his careful reading of this manuscript and useful comments.


\def\theequation{A. \arabic{equation}}
\setcounter{equation}{0}




\end{document}